# THE SCIENCE TRAINING PROGRAM FOR YOUNG ITALIAN PHYSICISTS AND ENGINEERS AT FERMILAB

Emanuela Barzi[1], Giorgio Bellettini[2], Simone Donati[2], Carmela Luongo[2]

[1]*Fermi National Accelerator Laboratory (US) and Ohio State University (US)*
[2]*University of Pisa and INFN (Italy)*

## Abstract

The summer training program for Italian undergraduate and graduate students at the Department of Energy (DOE) laboratory of Fermilab (Batavia, Illinois, USA) started in 1984 as a 2-month training program for Italian undergraduate students in physics of the Istituto Nazionale di Fisica Nucleare (INFN) collaborating in the Collider Detector experiment (CDF) at the Fermilab Tevatron proton-antiproton collider. While in 1984 the program involved only 4 physics students from the University of Pisa, in the following years it rapidly grew in scope and size under the management of the Cultural Association of Italians at Fermilab (CAIF). With an average number of 30 trainees/year reached in the last few years, the total number of Italian students hosted at Fermilab since 1984 has exceeded 500 units. Since 2015 the program has been included in the portfolio of the summer courses of the University of Pisa, which acknowledges 6 ECTS to the interns.

Besides describing several recent initiatives, which have expanded and strengthened the program, detailed information is given on the student recruiting process, on the training programs and on how the trainees' performance is evaluated.

## 1   FOREWORD

In this paper we review the past and describe the present training programs of Italian science students and young graduates at the Fermi National Accelerator Laboratory of the United States. The program started in 1984 with 4 undergraduate physics students from the University of Pisa. Now, more than thirty years later, each year we are managing on average 30 trainees in physics and engineering, who are distributed all over Fermilab as well as at some US Space Science laboratories.

The program is now part of the extensive Fermilab educational effort and is a summer course of the University of Pisa, which acknowledges 6 ECTS to the interns. It is run by CAIF under the sponsorship of INFN and of the Italian Embassy in Washington. We are looking forward to gain support from the Italian Minister of University and Research, and from new sponsors.

The success of our endeavour may suggest to colleagues interested in science education to repeat this experience elsewhere.

## 2   HISTORICAL BACKGROUND

Since 1984 the Italian groups of the Istituto Nazionale di Fisica Nucleare (INFN, http://home.infn.it/it/) performing experiments at the national DOE laboratory of Fermilab (Batavia, Illinois) have been running a 2-month summer training program for Italian students. For many years the program was motivated by the interest of the Italian groups of the Collider Detector (CDF, https://www-cdf.fnal.gov/) in raising students who would join their experiment. The CDF Italians could save enough money from their research funds for offering for two months to a few students living conditions comparable to those of the students of the CERN summer training program. Under the management of the Cultural Association of Italians at Fermilab (CAIF, http://caif.fnal.gov/), support was provided also by the American CDF groups who hosted physics students of the Italian universities where a CDF group was active, i.e. Pisa, Padua, Rome, Trieste, Udine, Bologna.

After completion of the upgraded CDF detector in 2004, the CDF budget was significantly reduced by INFN. On the other hand, more and more skilled students were applying and would have been very welcome in the program. The Italian groups could no longer afford covering the cost. As a result, the program was getting to a dead end. However, INFN and DOE intervened and signed an exchange agreement according to which about 4 students would be supported yearly by DOE at Fermilab, while INFN would provide the same support to American students visiting INFN laboratories in Italy. Since then this support has been granted every year and has become the solid root around which a much wider program could grow. In 2007 an agreement was reached between the School of Advanced Studies of Pisa (SSSA, http://www.sssup.it/) and Fermilab for sharing yearly the support for 4 engineering students of the School. In 2009 the Italian Scientists and Scholars in North America Foundation (ISSNAF, http://www.issnaf.org/) was able to raise funds from Italian sponsors, for training students primarily in astrophysics and space research institutions (INAF, www.inaf.it and ASI, www.asi.it). ISSNAF transferred funds to CAIF as needed to support training programs in the fields of interest of the sponsors. Accordingly, since then some students are being sent by CAIF to other labs for training in areas not covered by the Fermilab mission (see Appendix A).

At all times the Fermilab groups have contributed to supporting their Italian students, thus allowing expansion of the program beyond the limited DOE+SSSA+ISSNAF funds. Since the CDF Collaboration ended data-taking in 2011 and the physics analysis is rapidly getting to an end, within the Fermilab educational program the students are now assigned by CAIF to a wide spectrum of research and technical groups. Because of the several new experiments under construction at Fermilab, presently more engineering than physics students are requested by the groups.

## 3  RECRUITMENT

### 3.1  Students

In January each year we announce the available grants with a number of posters and flyers, slightly different from one another depending on the sponsor, whereby some basic information is given to the applicants. CAIF members and friends take care of appending posters in a number of Italian Universities. As of now we have not been able to get this information distributed c/o the Italian Minister of University and Scientific Research. We shall continue shooting for this goal, which would allow for a much more efficient transfer of information to all universities.

The list of qualified applicants is limited to physics students following the last year course of the triennial Laurea or the "Laurea Magistrale" courses (equivalent to an American Master degree). Similarly, the engineering students must be following their own Laurea Magistrale courses. The list of passed exams should be provided, and an average Italian grade higher than 26/30 is requested.

The program is international. Non-Italian students are accepted within rules that define equivalent qualifications as for Italians. Computer science skills and good knowledge of English must be stated in the application, to be later checked in the interview. Candidates are informed that they will have to discuss their case in person or by a Skype meeting with representatives of CAIF and of the sponsors. Thorough interviews are performed with great care by a board chaired by S. Donati. The profiles of the qualified students are made available to the potential supervisors who indicate those best suited for their program. Those for whom a good fit and work program are found within a Fermilab team will enter the USA with a J1 visa supported by the laboratory. Free housing and rental cars are provided besides a weekly salary, and some sponsors also cover the round-trip journey to the U.S. and the health insurance. Their salary, however, is sufficient to allow even students who pay for their travel to cover all costs encountered in two months of stay.

In early spring CAIF members explore the available training programs at the lab, to be matched to the skills and interests of the trainees. Agreement is also found with outside labs for training students sponsored by Space Science agencies. In late spring the winners are selected. As a rule, their training period is from the end of July to the end of September.

### 3.2  Young Engineers

CAIF hosted in the fall 2013 and 2014 a number of young professional engineers, proposed by the Italian Engineering Order (CNI, "Consiglio Nazionale degli Ingegneri", https://www.tuttoingegnere.it/PortaleCNI/), for internships on advanced technologies in USA. In 2014 CNI granted ISSNAF with 22 two-month fellowships in various U.S. labs, out of which 4 were assigned to CAIF. During the year, we have asked the Fermilab groups, all-over Particle Physics (PPD), Accelerator (AD), Computing (CD) and Technical Divisions (TD), to propose training programs. 23 programs were offered to CNI. Their impressive list is given in Appendix B. Many young engineers applied to these programs, but only 4 could be funded by CNI. The winners were selected by CAIF according to their profiles. The 23 proposed programs indicate to potential sponsors the outstanding role offered by Fermilab as a tutoring Institution in science and technology. The 4 assigned positions have been highlighted.

## 3.3 Laurea Students

Based on the strong interest that a number of former Italian summer students showed to come back to the lab and continue their research, E. Barzi started in 1998 what later became an official thesis program for a Laurea Magistrale. Some students were invited back to the lab and employed again for a period of about 6 months, as needed for performing an adequate research program for their "Specialized Laurea". Since 2009, she has been officially in charge of a TD Laurea Committee that assigns these contracts in the Technical Division. Although the funds made available have been rather limited and the program was confined to that Division, 23 Italian engineering students, out of which 9 (highlighted in Appendix C) had been previously summer students, have been supported by the lab for their Laurea program. A number of them are now permanent employees. This shows how the laboratory had a significant return from this program.

In the period 2003-2011, 22 students have obtained their three-annual "Short Laurea" with the work done during their two-month summer internship. Engineering students have been supported for longer stays by the lab, and many physics summer students got extended support by the INFN groups for furthering their studies at the lab. Out of about 200 Italian physics students who got their Specialized Laurea and PhD within the CDF Italian research groups, about 50% have been initially summer students. In the interest of the Italian students as well as of the Laboratory [4], CAIF is now supporting the proposal for a substantial increase in funding for the lab Laurea program.

## 4 LOGISTICS

A key action item is contacting research groups and finding out which ones are interested in offering a training program to Italian students. It is essential to find a match between the profiles of the candidates and the interest of the groups. Only students matching an available training program are ultimately selected. In order to reach a consistent picture, in springs CAIF sends the list of the qualified candidates to the supervisors and in turn informs the students of the available programs. By exchanging questions and information between students and supervisors, the list of accepted students can be created. The Fermilab Personnel Office is informed and e-mails are sent to the winners with the job offers needed to get the appropriate J1 entry visa.

Once the offer is accepted, the students are responsible for requesting a J1 entry visa to an U.S. Consulate in Italy. In addition, substantial paperwork is requested to the students by the Fermilab Visa Office. This includes accessing the FermiWorks website to complete an Onboarding process and filling a New Hire document. The rules for accessing the FERMI computing domain can also be followed from a distance in order to get access to the domain immediately upon arrival at the lab.

Upon arrival to the Chicago airport, students are instructed to get a limo to the lab, where they can pick up rental car and are instructed on how to reach their accommodation. CAIF Members are around watching and making sure that the process proceeds smoothly.

On their first workday at the lab the students are convened for an Orientation Session, where they are asked to provide proof of admission to USA and show a medical insurance. They are instructed on practical life rules in the U.S., like traffic rules, how to open a bank account and get a Social Security number, and additionally on specific Fermilab rules. They are introduced to their supervisors who are responsible for assigning them adequate office space and access to computers. Finally, they are instructed on how to get their work certified weekly by their supervisor as a condition for being timely paid.

A final session must be attended shortly before departure at the end of the training period. There students are instructed on how to handle their bank account at later times, how to fulfil U.S. tax regulations and how to adhere to the J1 visa conditions when coming back to the U.S.

Although this bureaucracy may look alarming, experience shows that students are able to learn quickly and fulfil all requests, and that in few days they are able to work efficiently.

Since at the end of July not enough laboratory dorms may be available to accommodate all students, some may be initially housed in nearby hotels. Within a few weeks, however, all students return to lab dorms. Accommodation on the lab site is very convenient for saving time, working more efficiently and living comfortably at the same time.

## 5 WORK PROGRAMS

The training programs span a very wide range of science and technology. The programs assigned in 2017 are listed in Appendix A. It is expected that they will be similar in 2018 For physicists, they include analysis of experimental data, set-up particle detectors, test of particle accelerator components. Programs for engineers include design of fast digital electronics, of detectors and of accelerator components, test of superconducting materials and magnets, high precision mechanics, advanced computing. Students make use of advanced computation means and programming languages, as C, C++, and Java, and apply advanced CAD and technical tools for mechanics and electronics design (MatLab, OrCAD, etc.). Physicists develop significant knowledge in statistical data analysis (Root). This work is performed within projects, by analysing data from experiments like CDF, Nova, Mu2e, DESI, Muon g-2, MicroBooNE, SBND, ICARUS, General Accelerator R&D and CMS.

The student is integrated as much as possible in his or her research group and is encouraged to interact with as many colleagues as possible, well beyond their supervisor. The supervisor should meet his or her student personally at least once per week in order to ensure the best productivity of the trainee. The students also participate in group meetings, where they present and discuss their results in a wide professional environment.

All students are requested to give a mid-term and a final oral presentation, and to write a technical report at the end of their stay. These documents are stored in the Education Office web archive. They can be easily accessed at http://eddata.fnal.gov/lasso/summerstudents/view.lasso and can be consulted to illustrate the excellence of the program

### 5.1 Internships funded by the Italian Space Agency (ASI) and National Institute of Astrophysics (INAF)

In 2010 ASI and INAF started providing financial support to CAIF for 2-month internships in US space science laboratories, similarly to the Fermilab program. Students' selection is made by CAIF members in collaboration with INAF or ASI personnel. As of now, host institutions in US have included the NASA Goddard Space Flight Center, the Space Telescope Science Institute in Baltimore, the Harvard Smithsonian Center for Astrophyscs, the NaSA Jet Propulsion Laboratory in Pasadena, Stanford University, Columbia University in New York, the University of Arizona in Tucson, the SLAC National Accelerator Laboratory, the University of Colorado at Boulder, Purdue University, and the University of Texas at Arlington.

## 6 GROWING INVOLVEMENT AT EUROPEAN LEVEL

### 6.1 The University of Pisa Summer School

In 2015 an ad-hoc summer course was approved at the University of Pisa to provide an academic framework to the Fermilab program. Interns are enrolled for the 9-week duration of the internship. They have a regular position as local students (i.e. Summer help), with a proper identification and insurance coverage. Upon successful completion of the internship with an accurate final report and an oral interview, they are acknowledged 6 ECTS in their Diploma Supplement (https://www.unipi.it/summerschool).

### 6.2 The University of Oxford "MovingKnowledge17" International Neutrino Summer Student Program

In 2017 the University of Oxford promoted "MovingKnowledge17", i.e. their first International Neutrino Summer Student Program, organized in collaboration with the University of Pisa and CAIF. The 3 Fermilab interns selected for the MovingKnowledge17 school spent 4 weeks in Oxford followed by the 9 weeks at Fermilab, performing research in the field of neutrino physics, with the MicroBooNE and Short Baseline Near Detector (SBND) experiments. It is expected that the program will be similar also in 2018.

## 6.3 Outreach effort of EU projects

In 2016 the Fermilab internship program has become part of the outreach activities of two European Projects, MUSE "Muon Campus in US and Europe contribution" (H2020-MSCA-RISE-2015, GA 690835, muse.lnf.infn.it), and NEWS "NEw WindowS on the Universe and technological advancements from trilateral EU-US-Japan collaboration" (H2020-MSCA-RISE-2016, GA 734303, risenews.df.unipi.it). INFN coordinates both European Projects, which count more than 15 universities and research institutes and 5 private companies in Europe, and 5 universities and research institutes in US (Fermilab, Caltech and Stanford University) and Japan (the National Astronomical Observatory of Japan, and the Institute of Cosmic Ray Research at the University of Tokyo). The EU provides financial support for the mobility of European researchers to the US and of Japan participants. Researchers involved in Muse and News have organized seminars and short courses on the topics of their research activities for the Fermilab interns. The 2016 seminars were mostly dedicated to the Muon (g-2) experiment, the 2017 seminars to the Mu2e experiment.

## 7 ENDORSEMENTS AND SPONSORSHIPS

The high quality of the CAIF student training programs was acknowledged in 2012 by INFN with the assignment of the Institute logo and with an annual donation. In 2013 the Italian Washington Embassy also authorized CAIF to make use its logo and requested CAIF to be sponsored by the Italian Minister of University and Scientific Research. In 2016 also ASI assigned its logo and an annual donation.

## 8 SUMMARY

A voluntary, self-managed education program started long ago by a few Italian physicists engaged in a particle physics experiment in the U.S. has turned into a solid, multi-disciplinary program bringing dozens of bright young undergraduate and graduate physicists and engineers to learn and contribute in high-tech research in the U.S. every year. Over the thirty-five years of its life, more than 500 students have taken part in the program. Table 1 reports the statistics relative to the programs of the last 10 years. Stimulated by the enthusiastic reports of their students, Italian universities now convey a strong feeling of appreciation for this program.

| Year | Accepted | INFN & FNAL groups | SSSA | ASI | INAF | CNI | No. Physicists | No. Engineers |
|---|---|---|---|---|---|---|---|---|
| 2008 | 20 | 14 | 6 | | | | 12 | 8 |
| 2009 | 22 | 18 | 4 | | | | 9 | 13 |
| 2010 | 24 | 18 | 2 | 2 | 2 | | 16 | 8 |
| 2011 | 23 | 17 | 4 | | 2 | | 16 | 7 |
| 2012 | 21 | 14 | 5 | 2 | | | 10 | 11 |
| 2013 | 23 | 16 | 4 | 2 | | 1 | 11 | 12 |
| 2014 | 25 | 15 | 4 | 2 | | 4 | 9 | 16 |
| 2015 | 35 | 28 | 3 | 3 | | 1 | 17 | 18 |
| 2016 | 40 | 33 | 4 | 3 | | | 21 | 19 |
| 2017 | 30 | 24 | 3 | 3 | | | 15 | 15 |

TABLE 1: Students statistics in the years 2008-2017. Numbers of accepted students are separated correspondingly to the sources of financial support (INFN is the Italian National Institute of Nuclear Physics, SSSA is the Sant'Anna School of Advanced Studies of Pisa, ASI is the Italian Space Agency, INAF is the Italian National Institute of Astrophysics, CNI is the Italian National Council of Engineering).

# 9 ACKNOWLEDGEMENTS

Many colleagues at Fermilab and elsewhere have contributed to the success of this program, by supervising the students and guiding them to take full advantage of the lab's capabilities. Their role has been and will be vital. The program would not have been possible without the strong support of the Lab Chief Operating Officer Timothy Meyer, and the highly professional and friendly help of the Visa and the Education Offices. We are grateful to INFN and to ASI for their support to CAIF, which provides the means for a vital and active Association.

This work was supported by the EU Horizon 2020 Research and Innovation Programme under the Marie Sklodowska-Curie Grant Agreement No. 734303.

# 10 APPENDIX A – SUMMER STUDENT TRAINING PROGRAMS IN 2017

## 10.1 Engineering Programs

1) Alessandro Giannetti – Mechanical Engineering, University of Pisa, "Facilitating Mu2e integration via 3D CAD model", Supervisor George Ginther, Technical Division – Fermilab.

2) Federico Nesti – Robotics and Automation Engineering, University of Pisa, "Vibrating wire control system for Mu2e", Supervisor Thomas Strauss, Technical Division – Fermilab.

3) Pietro Guardati – Electronics and Robotics Engineering, University of Pisa, "Automation and calibration magnet's plant for Mu2e", Supervisor Luciano Elementi, Technical Division – Fermilab.

4) Antonio Di Bello – Computer Engineering, Polytechnics of Milan, "Predictive model for Optimization of Grid Utilization", Supervisor Michael Kirby, Scientific Computing Division – Fermilab.

5) Mariano Basile – Computer Engineering, University of Pisa, "Migration of GRATIA active archive project accounting to GRACC", Supervisor Kevin Retzke, Scientific Computing Division – Fermilab.

6) Claudio Franceschelli – Mechanical Engineering, University of Pisa, "Sub-modelling of $Nb_3Sn$ 10-stack compression in elasto-plastic regime, and how to apply results to coils", Supervisors Alexander Zlobin and Emanuela Barzi, Technical Division – Fermilab.

7) Marco Castellari – Electronics and Telecommunication Engineering, University of Ferrara, "Novel RF tuning scheme for jacketed multi-cell SRF cavities using pressurized balloons for PIP-II", Supervisors Donato Passarelli and Mohamed Hassan, Technical Division – Fermilab.

8) Stefano Trivini – Material Science, University of Padova, "Combustion synthesis and characterization of gamma/delta-NbN for SRF cavity application and vortices study in superconductors", Supervisor Mattia Checchin, Technical Division – Fermilab.

9) Alessio Durante – Electrical Engineering, Polytechnics of Milan, "SILVACO simulation of a monolithic active pixel low energy sensor for X-rays detection in imaging experiment at LCLS-II", Supervisor Grzegorz Deptuch, Particle Physics Division – Fermilab.

10) Matteo Quagliotto – Energy Engineering, University of Pisa and Sant'Anna School of Advanced Studies of Pisa, "COMSOL simulations for heavy assembly building at IARC", Supervisor Charlie Cooper, Illinois Accelerator Research Center – Fermilab.

11) Paolo Vecchiolla – Energy Engineering, University of Pisa and Sant'Anna School of Advanced Studies of Pisa, "Current leads for SRF cavities cryomodule", Supervisor Vincent Roger, Technical Division – Fermilab.

12) Tommaso Rizzo – Electronics Engineering, University of Pisa and Sant'Anna School of Advanced Studies of Pisa, "Thermal chamber controller for burn-in system of CMS silicon modules", Supervisors Anadi Canepa and Lorenzo Uplegger, Particle Physics Division – Fermilab.

13) Paolo Cappuccio – Aerospace Engineering, University of Rome La Sapienza, "3D pendulum C++ model to simulate fuel slosh", Supervisor Hanspeter Schaub, Autonomous Vehicle Simulation (AVS) Laboratory, University of Colorado at Boulder.

14) Ernesto Poccia – Aerospace Engineering, University of Pisa and Sant'Anna School of Advanced Studies of Pisa, "Deterministic sampling-based algorithms for motion planning under differential constraints", Supervisor Marco Pavone, Autonomous Systems Laboratory, Department of Aeronautics and Astronautics, Stanford University.

15) Pasquale Walter Agostinelli, Aerospace Engineering, Polytechnics of Milan, "Filtered Rayleigh Scattering", Supervisor Luca Maddalena, Aerodynamics Research Center, University of Texas at Arlington.

## 10.2 Physics Programs

1) Stefano Falletta - Physics of Complex Systems, Alta Scuola Politecnica, Turin, "Electrochemical Nb3Sn film technologies", Supervisor Emanuela Barzi, Technical Division – Fermilab.

2) Francesco Lucarelli – Physics, University of Pisa, "Antiproton background rejection in the Mu2e experiment", Supervisor Robert Bernstein, Particle Phyisics Division – Fermilab.

3) Chiara Grieco – Physics, University of Naples Federico II, "Mu2e straw tube characterization with strontium source, and study of preamplifier-straw connection", Supervisor Manolis Kargiantoulakis, Particle Physics Division – Fermilab.

4) Paolo Girotti – Physics, University of Bologna, "The Muon g-2 experiment: Laser Calibration System", Supervisors Marco Incagli (INFN) and Brendan Casey, Particle Physics Division – Fermilab.

5) Claudio Andrea Manzari – Physics, University of Bari, "Search for Z' – pairs production decaying into Dark Matter and boosted Jets at CMS", Supervisor Matteo Cremonesi, Particle Physics Division – Fermilab.

6) Andrea Serafini - Physics, University of Ferrara, "Preliminary Study for LAr-TPC systematic errors", Supervisor Angela Fava, Neutrino Division – Fermilab.

7) Federico Roccati - Physics, Ludwig Maximilian University, Munich, "Final development and testing of the DAQ system for the ICARUS experiment", Supervisor Wesley Ketchum, Neutrino Division – Fermilab.

8) Matteo Mazzanti - Physics, University of Ferrara, "GENIE parameters tuning", Supervisors Minerba Betancourt and Patrick Stowell, Neutrino Division – Fermilab.

9) Giovanni Volta - Physics, University of Ferrara, "Physical interpretation of cluster pressure profile in CMB observation", Supervisor James Annis, Scientific Computing Division – Fermilab.

10) Bianca Giaccone - Physics, University of Milan, "Study of premature quench fields of nitrogen-doped niobium cavities", Supervisor Martina Martinello, Technical Division – Fermilab.

11) Livio Verra – Physics, University of Milan, "MC6 – MC7 simulation for Time-of-Flight LAPPD detectors", Supervisors Henry Frisch (University of Chicago) and Mandy Rominsky, Particle Physics Division – Fermilab.

12) Eleonora Pasino – Physics, University of Milan, "Search for high transverse momentum Higgs decaying in a pair of bottom quarks with CMS", Supervisor Caterina Vernieri, Particle Physics Division – Fermilab.

13) Michele Piero Blago – Physics, Ruprecht-Karls-Universitat, Heidelberg, "Cosmic ray studies for SBND", Supervisor Roxanne Guenette (Harvard University) and Corey Adams (Harvard University), Neutrino Division – Fermilab.

14) Karolina Rozwadowska – Physics, University of Warsaw, "Study of muon neutrino interactions in MicroBooNE", Supervisors Roxanne Guenette (Harvard University) and Marco del Tutto (Oxford University), Particle Physics Division – Fermilab.

15) Monika Venckauskaite – Physics, University of Vilnius, "Study of electromagnetic showers and cosmic-ray induced electromagnetic background in MicroBooNE", Supervisors Roxanne Guenette (Harvard University), Stefano Roberto Soleti (Oxford University) and Corey Adams (Harvard University), Neutrino Division – Fermilab.

## 11 APPENDIX B – AVAILABLE INTERNSHIPS FOR CNI ENGINEERS IN 2014

(only a few programs were actually assigned)

1) **ASSIGNED** to Claudio Pontili, Computing Eng. – Development of advanced distributed GRID/CLOUD computing, Gabriele Garzoglio, Steven Timm supervisors, Computing Division. One opening for a computing engineer in fall (best option) and/or in summer 2014.

2) Development of low-noise electronics for astrophysics experiments, Gustavo Cancelo supervisor, Electronic Department of the Computing Division. One opening for an electronic engineer preferably in fall 2014.

3) Creating and integrating new methods for accessing and delivering large data sets for distributed physics analysis. Andrew Norman, Marc Mengel supervisors, Scientific Data Processing Department of the Computing Division. One opening for a computing engineer in summer and/or in fall 2014.

4) Work in the team of software professionals of the Fermilab Computing Center of the CMS experiment at the LHC of CERN, to develop distributed monitoring of on-going data analysis in the world-wide grid and to improve the processing infrastructures of the Center. Oliver Gutsche, Dave Mason supervisors, Scientific Data Processing Department of the Computing Division. One opening for a computing engineer in summer and/or in fall 2014.

5) Evaluation of the mechanical accuracy achievable in the muon transport system of the Mu2e experiment, Rodger Bossert supervisor, Magnet System Department of the Technical Division. One opening for a mechanical engineer in summer 2014.

6) Simulation and construction of a prototype $Nb_3Sn$ superconducting helical solenoid for the Mu2e experiment, Mauricio Lopes supervisor, Magnet System Department of the Technical Division. One opening for a mechanical engineer in summer 2014.

7) Electrical modelling of toroid magnet coils and measurement and analysis of performances of dipole magnet prototypes for the Mu2e experiment at Fermilab, Luciano Elementi supervisor, Magnet Systems Department of the Technical Division. One opening for an electronic engineer in summer 2014.

8) Development of strain-resistant $Nb_3Sn$ superconducting strand and cables for applications in high field magnets, and electro-mechanical modelling of composite and anisotropic materials in the plastic regime, Emanuela Barzi supervisor, Superconducting Strand and Cable Group of the Technical Division. One opening for a mechanical engineer in summer and/or in fall 2014.

9) Construction of a ultra-high frequency spectrometer based on Caesium Iodide (CsI) to measure the Tera-Hz spectrum of a pico-second-electron beam, Vic Scarpine, Randy Thurman-Keup, Jayakar Thangaraj supervisors, Beam Instrumentation Department of the Accelerator Division. One opening for a mechanical/optical engineer in summer 2014.

10) Ceramic beam tube resistive coating R&D, Linda Valerio supervisor, Mechanical Support Department of the Accelerator Division. One opening for a Mechanical Engineer and/ or a Material Scientist in summer (preferred) or in fall 2014.

11) Construction and commissioning of a Titanium-sapphire multi-pass amplifier of a laser source, for studies of laser-induced particle beams acceleration, Philippe Piot supervisor, Experimental Beam Physics Department of the Accelerator Physics Center. One opening for a photonics-engineering or/and electrical-engineering in fall (preferred) or in summer 2014.

12) Simulation of the electron beam transport through the linac of the Advanced Superconducting Test accelerator ASTA of the Accelerator Physics Center, Jayakar Thangaraj supervisor. One opening for a computing engineer in summer 2014.

13) **ASSIGNED** to Giorgio Fasce, Electrical Eng. – Design of a gas-filled RF cavity for large-acceptance muon beam ionization cooling, Katsuya Yonehara supervisor, Muon Accelerator Department of the Accelerator Physics Center. One opening for an electrical or electronic engineer in summer or in fall 2014.

14) Development and production of materials for applications in improved scintillation detection, Anna Pla-Dalmau supervisor, Detector Development and Fabrication Department of the Particle Physics Division. One opening for a chemical or materials engineer in fall 2014.

15) Development of an increased sensitivity photo-detector for an optimized veto of cosmic ray muons for the Mu2e experiment, Paul Rubinov supervisor, Electrical Engineering Department of the Particle Physics Division. One opening for an electronic engineer in summer or in fall 2014.

16) Development of warm electronics for the CDMS detectors, Sten Hansen supervisor, Electrical Engineering Department of the Particle Physics Division. One opening for an electronic engineer in fall 2014.

17) Tests of three-dimensional integrated circuits bonded to pixelated sensors, Grzegorz Deptuch, Ron Lipton supervisors, Electrical Engineering Department of the Particle Physics division. One opening for an electrical engineer in summer or in fall 2014.

18) Testing of Application Specific Integrated Circuits (ASIC) bonded to high sensitivity single photon detection sensors, Farah Fahim supervisor, Electrical Engineering Department of the Particle Physics division. One opening for an electronic engineer in fall 2014.

19) Design of the cooling system of the upgraded cryogenic distillation column to separate Argon form an Argon/Nitrogen mixture, David Montanari supervisor, Mechanical Engineering Department of the Particle Physics Division. One opening for a cryogenic/thermal engineer in summer 2014.

20) Design of new gas distribution system for the upgraded cryogenic distillation column at PAB to separate Argon from an Argon/Nitrogen mixture, David Montanari supervisor, Mechanical Engineering Department of the Particle Physics Division. One opening for a mechanical engineer in summer 2014.

21) **ASSIGNED** to Martina Pagnani, Mechanical Eng. – Development of the mechanical structure for the calorimeter of the Mu2e experiment, Matteo Martini supervisor, Mu2e Group, Particle Physics Division. One opening for a mechanical engineer in summer or in fall 2014.

22) Development of the front-end electronics for the calorimeter of the Mu2e experiment, Fabio Happacher supervisor, Mu2e Group, Particle Physics Division. One opening for an electronic engineer in summer or in fall 2014.

23) **ASSIGNED** to Lisa Favilli, Civil Eng. – Design and construction of facilities and equipment to support the LBNE neutrino experiment at Fermilab and at the Sanford Underground Research Facility located in Lead, South Dakota (SURF), Tracy Lundin supervisor, LBNE Department of the Neutrino Division. One opening for a civil engineer in summer and/or in fall 2014.

## 12 APPENDIX C – LAB-SUPPORTED ENGINEERING LAUREAS SINCE 1999

1) Cristian Boffo, **former summer student** – Mechanical Eng., University of Udine, 1999: Magnetization Measurements at 4.2K of Multifilamentary Superconducting Strand, Prof. G. Pauletta, E. Barzi Advisors - TD (**originally hired at Fermilab**, later at Babcock Noell GmbH, Germany)

2) Michela Fratini – Nuclear Eng., Pisa University, 2002: A Device to Test Critical Current Sensitivity of $Nb_3Sn$ Cables to Transverse Pressure, Prof. C. Angelini, Prof. F. Fineschi, E. Barzi Advisors - TD (originally was offered a contract at CEA/Saclay, then hired at ENEL, Italy)

3) Sara Mattafirri – Nuclear Eng., Pisa University, 2002: Kinetics of Phase Growth during the Cu-Sn Diffusion Process and the $Nb_3Sn$ Formation. Optimization of Superconducting Properties, Prof. C. Angelini, Prof. F. Fineschi, Prof. S. Lanza, E. Barzi, J.M. Rey Advisors - TD (originally hired at LBNL, later at Nuovo Pignone, Florence)

4) Licia Del Frate – Nuclear Eng., Pisa University, 2004: Design of a Low Resistance Sample Holder for Instability Studies of Superconducting Wire, Prof. C. Angelini, Prof. F. Fineschi, Prof. S. Lanza, E. Barzi Advisors - TD (was offered a contract at Fermilab, then hired as a Ph.D. student at Pisa University, now in industry, Italy)

5) Vito Lombardo – Computing/Automation Eng., Sant'Anna School of Advanced Studies of Pisa, 2007: Automation of Short Sample Facility for Critical Current and Low Field Instability Measurements of Superconducting Strands at Cryogenic Temperatures, Prof. M. Innocenti, Prof. L. Pollini, E. Barzi, D. Turrioni Advisors - TD (**hired at Fermilab**)

6) Marco Danuso, **former summer student** – Mechanical Eng., Sant' Sant'Anna School of Advanced Studies of Pisa, 2008: Parametric Analysis of Forces and Stresses in Superconducting Magnets Windings, Prof. M. Beghini, E. Barzi, A. Zlobin Advisors - TD (hired at Finmeccanica, Italy)

7) Gabriella Norcia, **former summer student** – Mechanical Eng., Pisa University, 2009: Design of Modular Test Facility for HTS Insert Coils, Prof. M. Beghini, E. Barzi Advisors - TD (hired at Ansaldo, Italy)

8) Giuseppe Gallo – Mechanical Eng., Pisa University, 2010: Mechanical Modeling of Superconducting Rutherford-type Cable Fabrication, Prof. M. Beghini, Prof. L. Bertini, E. Barzi Advisors - TD (**hired at Fermilab**)

9) Antonio Bartalesi – Mechanical Eng., Pisa University, 2010: Design of High Field Solenoids made of High Temperature Superconductors, Prof. M. Beghini, E. Barzi Advisors - TD (originally was offered a contract at Fermilab, then moved to CERN)

10) Alessandro Quadrelli, **former summer student** – Electrical Eng., Pisa University, 2010: Automated Control of the Tuning of Superconducting RF Cavities, Mirko Marracci, Franco Bedeschi Advisors (Warren Schappert supervisor) - TD

11) Matteo Scorrano, **former summer student** – Electronic Eng., Pisa University, 2010: Development of a Control System for Superconducting Cavities with Fast Tuners, Giovanni Pennelli, Franco Bedeschi Advisors (Youri Pischalnikov supervisor) - TD

12) Simone Moio, **former summer student** – Physics Eng., Turin Polytechnics, 2011: Effect of Subelement Size, RRR and Strand Size on Stability of RRP $Nb_3Sn$ Strands, Prof. R. Gonnelli, E. Barzi Advisors - TD (hired at Magneti Marelli, Italy)

13) Donato Passarelli – Mechanical Eng., Pisa University, 2011: Analysis of the mechanical behavior of the Superconducting Single Spoke Resonator type 1 in tests at cryogenic temperature, Prof. M. Beghini, L. Ristori Advisors - TD (**hired at Fermilab** and Ph.D. student at Pisa University)

14) Federico Puccinelli – Mechanical Eng., Pisa University, 2011: Detector support structure and installation system for the Mu2e experiment, Prof. Marco Beghini, Sandor Feher, Rodger Bossert Advisors - TD

15) Paolo Berrutti – Physics Eng., Turin Polytechnics, 2011: Radio frequency design and optimization of the low-medium-beta section for a 3 GeV linear particle accelerator, Prof. Gianni Coppa, Slava Yakovlev Advisors - TD (**hired at Fermilab**)

16) Andrea Pisoni – Physics Eng., Milan Polytechnics, 2012: Implementation of Nano Scale Formations in A15 Brittle Superconductors, Prof. R. Bertacco, E. Barzi, T. Rajh Advisors - TD (hired as a Ph.D. student at University of Lausanne)

17) Giulia Collura, **former summer student** - Electronic Eng., Turin Polytechnics, 2012: Beam Test of a High-Pressure RF Cavity for the Muon Collider, Prof. Felice Iazzi Advisor (Alvin Tollestrup Supervisor) - AD

18) Pietro Giannelli – Electronic Eng., Turin Polytechnics, 2012: Design of a Signal Conditioner for the Fermilab Magnet Test Facility, Prof. Marco Parvis Advisor (Darryl Orris Supervisor) - TD

19) Vincenzo Li Vigni – Electronic Eng., Palermo University, 2012: Design and testing of a digital regulator for Fermilab magnet power systems, Prof. Giuseppe Capponi, Valeria Boscaino, Andrzej Makulski Advisors - TD

20) Silvia Zorzetti, **former summer student** - Electronic Eng., Pisa University, 2013: Development of the world's first digital direct-current current transformer (DCCT) to measure particle beam intensities, Prof. Luca Fanucci, Manfred Wendt Advisors - AD (hired as a Ph.D. student at Pisa University/CERN, **presently Bardeen's Fellow at Fermilab**)

21) Matteo Grandini – Mechanical Eng., Pisa University, 2014: Design of a liquid helium transfer line support system, Prof. Bernardo Disma Monelli, Thomas Page Advisors - TD

22) Federico Reginato, **former summer student** – Physics Eng., Milan Polytechnics, 2014: Electrochemical Synthesis of Nb-Sn Coatings for High Field Accelerator Magnets, Prof. Silvia Franz, E. Barzi Advisors - TD

23) Andrea Palagi – Mechanical Eng., Pisa University, 2014: ASME Verification and Pressure Sensitivity optimization of the Cryostat of the 1.3 GHz Cavity, Prof. Marco Beghini, Prof. Leonardo Bertini, Allan Rowe Advisors – TD

24) Omar Al Atassi – Computing/Automation Eng., Pisa University, 2015: Inductance studies for the quench protection system of the Mu2e TS solenoids, Prof. Mario Innocenti Advisor (Darryl Orris Supervisor) – TD (**hired at Fermilab**)

25) Irene Nutini, **former summer student** – Physics, Firenze University, 2015: Study of charged particles interaction processes on Ar in the 0.2-2.0 GeV energy range through combined information from ionization free charge and scintillation light, Prof. Oscar Adriani and Flavio Cavanna Advisors – ND (presently a PhD student at the Gran Sasso Science Institute GSSI)

26) Marco Del Tutto, **former summer student** – Physics, La Sapienza University, Rome, 2015: Neutrino beam simulations and data checks for the NOVA experiment, Prof. Giovanni Rosa, Giulia Brunetti and John Cooper Advisors – ND (presently a PhD student at the University of Oxford)

27) Andrea Scarpelli – Physics, Ferrara University, 2016: Development of a synchrotron radiation beam monitor for the Integrable Optics Test Accelerator", Prof. Eleonora Luppi, Giulio Stancari Advisors - AD

28) Federica Bradascio, **former summer student** – Physics, University of Pisa, 2016: Studies of the impact of magnetic field uncertainties on physics parameters of the Mu2e experiment, Prof. Giorgio Bellettini, Konstantinos Vellidis Advisors – PPD (presently a PhD student at DESY).

29) Veronica Ilardi, **former summer student** – Mechanical Eng., University of Ferrara, 2016: Thermal analysis of the thermal shield and warm bore in the Mu2e transport solenoid, Prof. S. Piva, Tom Nicol, Mauricio Lopes Advisors – TD

30) Irene Zoi, **former summer student** – Physics, Firenze University, 2016: Characterization of radiation hard pixel sensors for the CMS Phase-2 upgrades, Prof. Raffaello D'Alessandro, Gino Bolla Advisors – PPD

31) Dante Totani, **former summer student** – Physics, L'Aquila University, 2016: Development of front-end electronics for a large surface SiPMs Array at cryogenic temperature, Profs. F. Villante and Flavio Cavanna Advisors - ND (presently a PhD student at University of L'Aquila and **Fermilab International Student**)

32) Giacomo Scanavini, **former summer student** – Physics, Pisa University, 2017: First Measurement of One Pion Production in Charged-Current Neutrino and Antineutrino Events on Argon, Prof. Alessandro Baldini and Ornella Palamara Advisors – ND (presently a PhD student at Yale University)

33) Franco di Ciocchis, **former summer student** – Mechanical Eng., Pisa University, in progress: Strategy and tools for assembling the SSR1 cold mass in its vacuum vessel for PIP-II, Prof. Ciro Santus, Donato Passarelli Advisors – TD (ongoing)